\newif\ifpdf
\ifx\pdfoutput\undefined
\pdffalse 
\else
\pdfoutput = 1 
\pdftrue
\fi

\documentclass [11pt, a4paper, openright, oneside] {article}

\usepackage {amsmath, amssymb, amsfonts,bbold,bm}
\usepackage {graphics}
\usepackage {graphicx}
\usepackage {longtable}
\usepackage {pgf,tikz}
\usepackage {float,afterpage} 
\usepackage {verbatim} 
\usetikzlibrary{arrows,automata,shapes,snakes,backgrounds,topaths,petri}
\usepackage {subfigure}
\usepackage{cite}

\title{Simulation-based model selection for dynamical systems in systems and population biology}
\author{Tina Toni\,$^{1,2,*}$ and Michael P. H. Stumpf\,$^{1,2}$\footnote{to whom correspondence should be addressed: ttoni@imperial.ac.uk, m.stumpf@imperial.ac.uk}\\\\
$^{1}$Division of Molecular Biosciences, $^{2}$Institute of Mathematical Sciences, \\
Imperial College London, UK}
\date{}

\begin{document}

\maketitle

Computer simulations have become an important tool across the biomedical sciences and beyond. For many important problems several different models or hypotheses exist and choosing which one best describes reality or observed data is not straightforward. We therefore require suitable statistical tools that allow us to choose rationally between different mechanistic models of e.g. signal transduction or gene regulation networks. This is particularly challenging in systems biology where only a small number of molecular species can be assayed at any given time and all measurements are subject to measurement uncertainty. Here we develop such a model selection framework based on approximate Bayesian computation and employing sequential Monte Carlo sampling. We show that our approach can be applied across a wide range of biological scenarios, and we illustrate its use on real data describing influenza dynamics and the JAK-STAT signalling pathway. Bayesian model selection strikes a balance between the complexity of the simulation models and their ability to describe observed data. The present approach enables us to employ the whole formal apparatus to any system that can be (efficiently) simulated, even when exact likelihoods are computationally intractable. 

\section{Introduction}

{M}athematical models are widely used to describe and analyze complex systems and processes.   Formulating a model to describe, e.g. a signalling pathway or host parasite system, requires us to condense our assumptions and knowledge into a single coherent framework \cite{May:2004p11952}. Mathematical analysis and computer simulations of such models then allow us to compare model predictions with experimental observations in order to test, and ultimately improve these models. The continuing success, for example of systems biology, relies on the judicious combination of experimental and theoretical lines of argument. 
\\\\
Because many of the mathematical models in biology (as in many other disciplines) are too complicated to be analyzed in a closed form, computer simulations have become the primary tool in the quantitative analysis of very large or complex biological systems. This, however, can complicate comparisons of different candidate models in light of (frequently sparse and noisy) observed data. Whenever probabilistic models exist, we can employ standard model selection approaches of either a frequentist, Bayesian, or information theoretic nature \cite{Burnham:2002p4089,Vyshemirsky:2008p14865}. But if suitable probability models do not exist, or if the evaluation of the likelihood is computationally intractable, then we have to base our assessment on the level of agreement between simulated and observed data. This is particularly challenging when the parameters of simulation models are not known but must be inferred from observed data as well. Bayesian model selection side-steps or overcomes this problem by marginalizing (that is integrating) over model parameters, thereby effectively treating all model parameters as nuisance parameters.
\\\\
For the case of parameter estimation when likelihoods are intractable, approximate Bayesian computation (ABC) frameworks have been applied successfully \cite{Beaumont:2002p13862,Marjoram:2003p5,Sisson:2007p2,Ratmann:2007p18882,Toni:2009p20998,Ratmann:2009p27795}. In ABC the calculation of the likelihood is replaced by a comparison between the observed data and simulated data. Given the prior distribution $P(\theta)$ of parameter $\theta$, the goal is to approximate the posterior distribution, $P(\theta|D_0) \propto f(D_0|\theta)P(\theta)$, where $f(D_0|\theta)$ is the likelihood of $\theta$ given the data $D_0$. ABC methods have the following generic form:
\begin{enumerate}

 	\item [\bf{1}] Sample a candidate parameter vector $\theta^*$ from prior distribution $P(\theta)$.

	\item [\bf{2}] Simulate a data set $D^*$ from the model described by a conditional probability distribution $f(D|\theta^*)$.

	\item [\bf{3}] Compare the simulated data set, $D^*$, to the experimental data, $D_0$, using a distance function, $d$, and tolerance $\epsilon$; if $d(D_0, D^*)
	\leq \epsilon$, accept $\theta^*$. The tolerance $\epsilon \geq 0$ is the desired level of agreement between  $D_0$ and $D^*$.
\end{enumerate}
The output of an ABC algorithm is a sample of parameters from the distribution $P(\theta|d(D_0, D^*) \leq \epsilon)$. If $\epsilon$ is sufficiently small then this distribution will be a good approximation for the ``true'' posterior distribution, $P(\theta|D_0)$. A tutorial on ABC methods is available in the Suppl. Material. 
\\\\
Such a parameter estimation approach can be used whenever the model is known. However, when several plausible candidate models are available we have a model selection problem, where both the model structure and parameters are unknown. In the Bayesian framework, model selection is closely related to parameter estimation, but the focus shifts onto the marginal posterior probability of model $m$ given data $D_0$,
$$
P(m|D_0) =\frac{ P(D_0|m)P(m)}{P(D_0)}
$$
where $P(D_0|m)$ is the marginal likelihood and $P(m)$ the prior probability of the model \cite{Gelman:2003}. This  framework has some conceptual advantages over classical hypothesis testing: for example, we can rank an arbitrary number of different non-nested models by their marginal probabilities;  and rather than only considering evidence against a model the Bayesian framework also weights evidence in a model's favour \cite{Jeffreys:1939p26845}. In practical applications, however, a range of potential pitfalls need considering: model probabilities can show strong dependence on model and parameter priors; and the computational effort needed to evaluate these posterior distributions can make these approaches cumbersome. 
\\\\
The computationally expensive step in Bayesian model selection is the evaluation of the marginal likelihood, which is obtained by marginalizing over model parameters; i.e. $P(D_0|m)=\int f(D_0|m,\theta)P(\theta|m)d\theta$, where $P(\theta|m)$ is the parameter prior for model $m$. Here we develop a computationally efficient ABC model selection formalism based on a sequential Monte Carlo (SMC) sampler. We show that  our ABC SMC procedure allows us to employ the whole paraphernalia of the Bayesian model selection formalism, and we illustrate the use and scope of our new approach in a range of models: chemical reaction dynamics, Gibbs random fields, and real data describing influenza spread and JAK-STAT signal transduction.

\section{ABC for model selection}

Our goal is to estimate the marginal posterior distribution of a model, $P(m|D_0)$, and in this section we explain two ways in which this problem can be approached. In the \textit{joint space based approach} we define a joint space of model indicators, $m=1,2,\ldots,|\mathcal{M}|$, and corresponding model parameters, $\theta$, obtain the joint posterior distribution over the combined space of models and parameters, $P(\theta,m|D_0)$, and finally marginalize over parameters to obtain $P(m|D_0)$. In the second, \textit{marginal likelihood based approach}, we estimate marginal likelihoods (also called \textit{the evidence}), $P(D_0|m)$, for each given model, and use these to calculate the marginal posterior model distributions through
$$
P(m|D_0) = \frac{P(D_0|m)P(m)}{\sum_{m'}P(D_0|m')P(m')}. 
$$
Both approaches have been applied under the ABC rejection scheme, which is computationally prohibitive for models with even an only moderate number of parameters \cite{Wilkinson:2007p14173, Grelaud:2009p29539}. Here we incorporate ideas from SMC to both of the above approaches, making them computationally more efficient. In this section we present only the more powerful approach \textit{ABC SMC model selection on the joint space}. We refer the reader to the Suppl. Material for derivations and details, as well as discussion on the ABC SMC model selection algorithm based on the marginal likelihood approach. 
\\\\
In model selection based on ABC rejection we adapt the basic ABC procedure (presented in the introduction) to the joint space, where \textit{particles} $(m,\theta)$ consist of a model indicator $m$ and a parameter $\theta$. The ABC rejection model selection algorithm on the joint space proceeds as follows \cite{Grelaud:2009p29539}:
\begin{enumerate}
	\item [\bf{1}] Draw $m^*$ from the prior $P(m)$.
	\item [\bf{2}] Sample $\theta^{*}$ from the prior $P(\theta|m^{*})$.
	\item [\bf{3}] Simulate a candidate data set $D^{*} \sim f(D|\theta^{*},m^{*})$.
	\item [\bf{4}] Compute the distance. If $d(D_0, D^*) \leq \epsilon$, accept $(m^*,\theta^*)$, otherwise reject it.
	\item [\bf{5}] Return to 1.
\end{enumerate}
Once a sample of $N$ particles has been accepted, the marginal posterior distribution is approximated by 
$$
P(m=m'|D_0) \approx \frac{\# \textrm{accepted particles} (m',.)}{N}. 
$$
In the ABC SMC model selection algorithm on the joint space, particles (parameter vectors)
$\{\! (m_1\!,\!\theta_1),\ldots,(m_N\!,\!\theta_N)\! \}$ are sampled from the prior distribution, $P(m,\theta)$, and propagated through a sequence of intermediate distributions, $P(m,\theta|d(D_0, D^*) \leq \epsilon_i)$, $i = 1,\ldots,T-1$,
until they represent a sample from the target distribution, $P(m,\theta|d(D_0, D^*)\leq \epsilon_T)$. The tolerances $\epsilon_i$ are chosen such that $\epsilon_1 > \ldots > \epsilon_T \geq 0$, and the distributions thus gradually evolve towards the target posterior distribution.
\\\\
The algorithm is presented below (and explained in the Suppl. Tutorial). 
\subsection*{ABC SMC model selection algorithm on the joint space}

\begin{enumerate}

 	\item [\bf{MS1}] Initialize $\epsilon_1, \ldots, \epsilon_T$.\\
	Set the population indicator $t = 1$. 

	\item [\bf{MS2.0}] Set the particle indicator $i = 1$.

	\item [\bf{MS2.1}] If $t=1$, sample $(m^{**},\theta^{**})$ from the prior distribution $P(m,\theta)$.\\
	If $t >1$, sample $m^{*}$ with probability $P_{t-1}(m^{*})$ and draw $m^{**}\sim KM_t (m|m^{*})$.\\
	Sample $\theta^{*}$ from previous population $\{\theta(m^{**})_{t-1}\}$ with weights $w_{t-1}$ and draw $\theta^{**}\sim KP_{t,m^{**}}(\theta|\theta^*)$. \\
	If $P(m^{**},\theta^{**}) = 0$, return to \textbf{MS2.1}.\\
	Simulate a candidate data set $D_{(b)} \sim f(D|m^{**},\theta^{**})$ $B_t$ times ($b=1,\ldots,B_t$) and calculate $b_t(m^{**},\theta^{**})$. \\
	If $b_t(m^{**},\theta^{**}) = 0$, return to \textbf{MS2.1}.

	\item [\bf{MS2.2}] Set $(m_t^{(i)},\theta_t^{(i)}) = (m^{**}, \theta^{**})$ and calculate the weight of the particle as
	$$
	w_t^{(i)}(m_t^{(i)},\theta_t^{(i)}) = 
	\left\{ \begin{array}{ll}
	b_t(m_t^{(i)},\theta_t^{(i)}),  &  \textrm{if } t=1  \\
	\normalsize{\dfrac
	{P(m_t^{(i)},\theta^{(i)}_t) b_t(m_t^{(i)},\theta_t^{(i)}) }
	{
	S
	}}
	, & \textrm{if } t >1.
	\end{array} \right.
	$$	 
	where
	\begin{eqnarray*}
	b_t(m_t^{(i)},\theta_t^{(i)})&=& \frac{1}{B_t}\sum_{b=1}^{B_t}\mathbb{1}(d(D_0, D_b^*)\le\epsilon_t) \\
	S &=& \sum_{j=1}^{|\mathcal{M}|} P_{t-1}(m_{t-1}^{(j)}) KM_t (m_t^{(i)}|m_{t-1}^{(j)}) \times \\
	&& \sum_{k;m_{t-1}=m_t^{(i)}} \frac{w_{t-1}^{(k)} KP_{t,m_t^{(i)}} (\theta_t^{(i)}|\theta_{t-1}^{(k)})}{P_{t-1}(m_{t-1}=m_t^{(i)})}
	\end{eqnarray*}
	If $i < N$ set $i = i+1$, go to \textbf{MS2.1}.  
	
	\item [\bf{MS3}] Normalize the weights $w_t$.\\
	Sum the particle weights to obtain marginal model probabilities,
	$$P_t(m_t=m) = \sum_{i;m^{(i)}_t=m} w_t^{(i)}(m_t^{(i)},\theta_t^{(i)}).$$
	If $t<T$, set $t = t+1$, go to \textbf{MS2.0}.   

\end{enumerate}
Particles sampled from a previous distribution are denoted by a single asterisk, and after perturbation  by a double asterisk. $KM$ is a model perturbation kernel which allows us to obtain model $m$ from model $m^*$ and $KP$ is the parameter perturbation kernel. $B_t \geq 1$ is the number of replicate simulation run for a fixed particle (for deterministic models $B_t = 1$) and $|\mathcal{M}|$ denotes the number of candidate models.
\\\\
The output of the algorithm, i.e. the set of particles $\{(m_T,\theta_T)\}$ associated with weights $w_T$, is the approximation of the full posterior distribution on the joint model and parameter space. The approximation of the marginal posterior distribution of the model obtained by marginalization is
$$
P_T(m_T=m) = \sum_{i;m_T^{(i)}=m} w_t^{(i)}(m_T^{(i)},\theta_T^{(i)}),
$$
and we can also straightforwardly obtain the marginalized parameter distributions.
\\\\
The algorithm requires the user to define the prior distribution, distance function, tolerance schedule and perturbation kernels. In all examples presented in the results section we choose uniform prior distributions for all parameters and models; that is all models are \textit{a priori} equally plausible. Such priors are informative in a sense that they define a feasible parameter region (e.g. reaction rates are positive), but they are predominantly non-informative as they do not specify any further preference for particular parameter values. This way the inference will mostly be informed by the information contained in the data. A good tolerance can be found empirically by trying to reach the lowest distance feasible and arrive at the posterior distribution in a computationally efficient way. Our perturbation kernels are component-wise truncated uniform or Gaussian and are automatically adapted by feeding back information on the obtained parameter ranges from the previous population. Distance functions are defined for each model as specified in the results section. The algorithm presented in Toni \textit{et al.} \cite{Toni:2009p20998} is a special case of the above algorithm for discrete uniform $KM$ kernel and uniform prior distribution of the model $P(m)$.

\section{Results}

In this section we illustrate ABC SMC for model selection on a simple example of stochastic reaction kinetics. We then compare the computational efficiency of ABC SMC for stochastic models of Gibbs random fields with that of the ABC rejection model selection method. Finally, we apply the algorithm to several real datasets: first we select between different stochastic models of influenza epidemics (where we can compare our approach with previously published results obtained using exact Bayesian model selection), and then apply our approach to choose from among different mechanistic models for the STAT5 signaling pathway.

\subsection{Chemical Reaction Kinetics}

\begin{figure}[thp]	
	\centering
	\includegraphics[width=8.7cm]{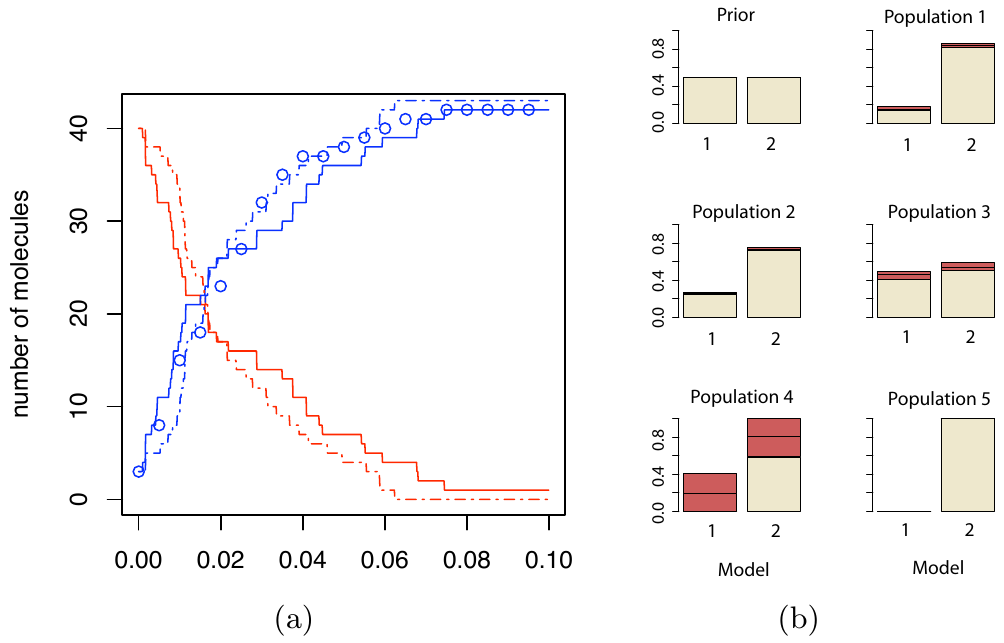} 
	\caption{\small{(a) Stochastic trajectories of species X (red) and Y (blue). Model 1 is simulated for $k_1=2.1$ (dashed line), model 2 for $k_2 = 30$ (full line). Data points are represented by circles. (b) We have repeated the model selection run 20 times; the red sections present 25\% and 75\% quantiles around the median. Prior distribution $P(m)$ is chosen uniform and $k_1$, $k_2$ $\sim U(0,100)$. Perturbation kernels are chosen as follows: $KP_t(k|k^*) = k^* + U(-\sigma,\sigma)$, $\sigma = 2 (\max\{k\}_{t-1} -\min\{k\}_{t-1}) $ and $KM_t(m|m^*) = 0.7$ if $m = m^*$ and $0.3$ otherwise. Number of particles $N=1000$. $B_t = 1$. Distance function is mean squared error and tolerance schedule $\epsilon = \{3000, 1400, 600, 140, 40\}$. \label{prion_figures}}}
\end{figure}

We illustrate our algorithm for the stochastic reaction kinetic models $X+Y \xrightarrow{k_1} 2Y$ and $X \xrightarrow{k_2} Y$. The first is a model of an autocatalytic reaction, where the reaction product Y is the catalyst for the reaction. In the second, molecules $Y$ do not need to be present for a change from $X$ to $Y$ to occur. Such models have, for example, been considered in the context of prion replication dynamics \cite{Prusiner:1982p26381, Eigen:1996p26567}, where $X$ represents a healthy form of a prion protein and $Y$ a diseased form.
\\
\\
We simulate synthetic datasets of $Y$ measured at 20 time points using Gillespie algorithm \cite{Gillespie:1977p13997} from model 2 with parameter $k_2 = 30$ and initial conditions  $X_0 = 40$, $Y_0 = 3$ (Figure \ref{prion_figures}(a), Suppl. Table 1). We apply our ABC SMC algorithm for model selection, which identifies the correct model with high confidence (Figure \ref{prion_figures}(b)).

\subsection{Gibbs random fields}
\begin{figure}[thp]	
	\centering
	\includegraphics[width=8.7cm]{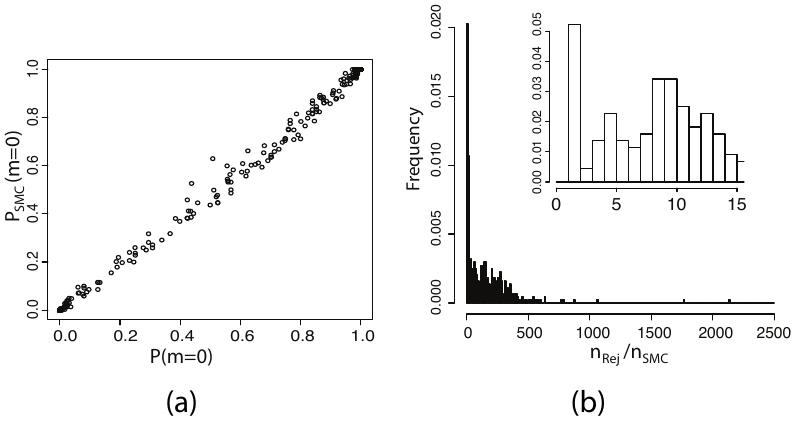} 
	\caption{\small{(a) True vs. inferred posterior model distribution. In ABC SMC we use the Euclidian distance
$d(D_0,x) = \sqrt{(S_0(D_0)-S_0(x))^2 + (S_1(D_0)-S_1(x))^2}$. $N=500$. $B_t = 1$. Tolerance schedule: $\epsilon = \{9,4,3,2,1,0\}$. Perturbation kernels: $KM_t(m|m^*) = 0.75$ if $m = m^*$ and $0.25$ otherwise; $KP_t(\theta|\theta^*) = \theta^* + U(-\sigma,\sigma)$, $\sigma = 0.5 (\max\{\theta\}_{t-1} -\min\{\theta\}_{t-1})$. We have excluded those datasets for which all states are in 0 or 1 (for  
which $P(m=0)\approx 0.3094$ is also correctly inferred) from the analysis. (b) Comparison of the number of simulation steps needed by ABC rejection ($n_{Rej}$) and ABC SMC ($n_{SMC}$); ABC SMC yields an approximately $50$-fold speed-up on average.
\label{GRF}}}
\end{figure}

Gibbs random fields have become staple models in machine learning, including applications in computational biology and bioinformatics (see for example \cite{Wei:2007,Grelaud:2009p29539}). Here we  use two Gibbs random field models \cite{Moller:2003p26909}, for which closed form posterior distributions are available. This allows us to compare the ABC SMC approximated posterior distributions of the models to true posterior distribtuions, and to demonstrate the computational efficiency of our approach when compared to model selection based on ABC rejection sampling. 
\\\\
Both models, $m_0$ and $m_1$, are defined on a sequence of $n$ binary random variables, $x = (x_1,\ldots,x_n)$, $x_i\in \{ 0,1 \}$; $m_0$ is a collection of $n$ iid Bernoulli random variables with probability $\theta_0/(1+\exp(\theta_0))$; $m_1$ is equivalent to a standard Ising model, i.e. $x_1$ is taken to be a binary random variable and $P(x_{i+1}=x_i|x_i)=\theta_1/(1+\exp(\theta_1))$ for $ i=2,\ldots,x_n$. The likelihood functions are
$$
f_0(x|\theta_0)\! =\! \frac{e^{\theta_0 S_0(x)}}{(1+e^{\theta_0})^n}\hspace{0.2cm}
\text{ and }\hspace{0.2cm}
f_1(x|\theta_1)\! =\! \frac{e^{\theta_1 S_1(x)}}{2(1+e^{\theta_1})^{n-1}},
$$
where $S_0(x) = \sum_{i=1}^{n} \mathbb{1}(x_i = 1)$ and $S_1(x) = \sum_{i=2}^{n} \mathbb{1}(x_i = x_{i-1})$ are sufficient statistics, respectively.
\\
\\
We simulate $1000$ datasets from both models for different values of parameters $\theta_0 \sim U(-5,5)$, $\theta_1 \sim U(0,6)$ and $n=100$. Using ABC SMC for model selection allows us to estimate posterior model distributions correctly and demonstrate a considerable computational speed-up in ABC SMC compared to ABC rejection (Figure \ref{GRF}).

\subsection{Infuenza infection outbreaks}

\begin{figure}[thp]	
	\centering
	\includegraphics[width=6cm]{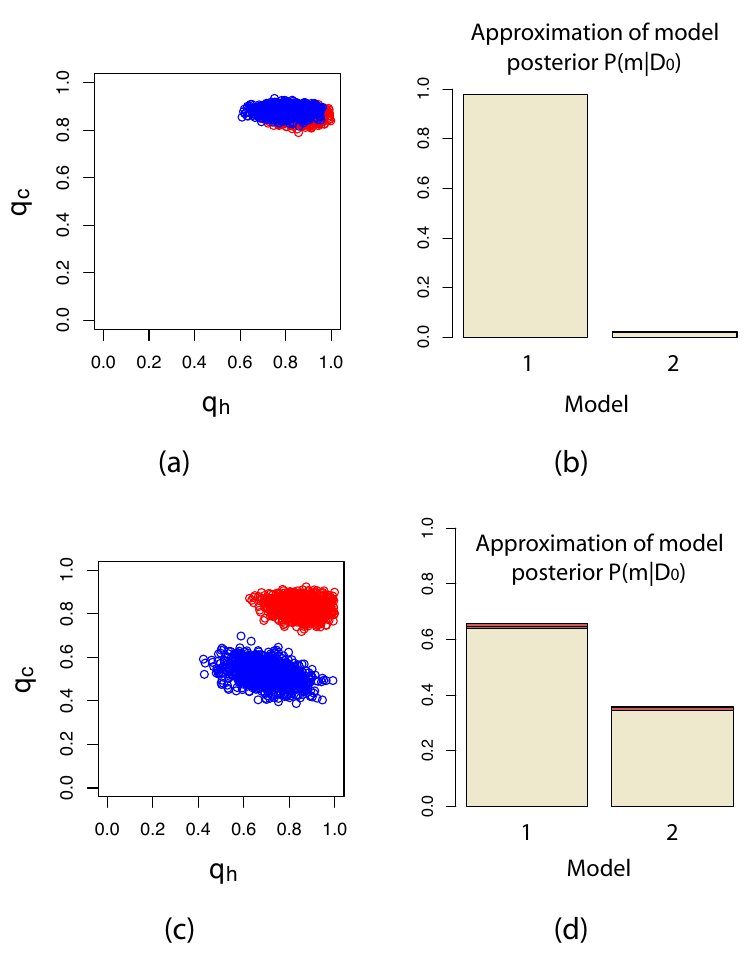}  
\caption{\small{(a) ABC SMC posterior distributions for parameters inferred for a four-parameter model from the data in Suppl. Table 2. Marginal posterior distributions of parameters $q_{c1}$, $q_{h1}$ (red) and $q_{c2}$, $q_{h2}$ (blue). 
(b) Estimation of a posterior marginal distribution $P(m|D_0)$. Model 1 is a two-parameter and model 2 a four-parameter model (\ref{basic}). 
All intermediate populations are shown in Suppl. Figure 1(a).
(c) The same as (a) but here the data used is from Suppl. Table 3. 
(d) Estimation of a posterior marginal distribution. Model 1 is a two-parameter and model 2 a three-parameter model (\ref{basic}). All intermediate populations are shown in Suppl. Figure 1(b).
\label{fig:influenza} }}
\end{figure}

We next apply ABC SMC for model selection to models of the spread of different strains of the influenza virus. We use  data from influenza A (H3N2) outbreaks that occurred in 1977-78 and 1980-81 in Tecomseh, Michigan \cite{Addy:1991p23540} (Suppl. Table 2), and a second dataset of an influenza B infection outbreak in 1975-76 and influenza A (H1N1) infection outbreak in 1978-79 in Seattle, Washington \cite{LonginiJr:1982p23531} (Suppl. Table 3). The basic questions to be addressed here are whether (i) different outbreaks of the same strain and (ii) outbreaks of different molecular strains of the influenza virus can be described by the same model of disease spread. 
\\\\
We assume that virus can spread from infected to susceptible individuals and distinguish between spread inside households or across the population at large \cite{LonginiJr:1982p23531}. 
Let $q_c$ denote the probability that a susceptible individual does not get infected from the community and $q_h$ the probability that a susceptible individual escapes infection within their household. Then $w_{js}$, the probability that $j$ out of the $s$ susceptibles in a  household become infected, is given by
\begin{equation}
w_{js} = \binom{s}{j}w_{jj}(q_c q_h^j)^{s-j}, \label{basic}
\end{equation}
where $w_{0s} = q_c^{s}$, $s=0,1,2,\ldots$, and
$
w_{jj} = 1-\sum_{i=0}^{j-1}w_{ij}. 
$
We are interested in inferring the pair of parameters $q_h$ and $q_c$ of the model (\ref{basic}) using the data from Suppl. Table 2. These data were obtained from two separate outbreaks of the same strain, H3N2, and the question of interest is whether these are characterized by the same epidemiological parameters (this question was previously considered in \cite{Clancy:2007p23340, ONeill:2000p23341}). To investigate this issue, we consider two models: one with four parameters, $q_{h1}$, $q_{c1}$, $q_{h2}$, $q_{c2}$, which describes the hypothesis that each outbreak has its own characteristics; the second  models the hypothesis that both outbreaks share the same epidemiological  parameter values for $q_h$ and $q_c$. Prior distributions of all parameters are chosen to be uniform over the range $[0,1]$.
\\
\\
To apply ABC SMC, we use a distance function
$$
d(D_0, D^*) = \frac{1}{2}(|| D_1 - D^*(q_{h1},q_{c1})||_{F} + || D_2 - D^*(q_{h2},q_{c2})||_{F}),
$$
where $||$ $||_F$ denotes the Frobenious norm, $D_0 = D_1 \cup D_2$ with $D_1$ the 1977-78 outbreak and $D_2$ the 1980-81 outbreak datasets from Suppl. Table 2, and $D^*$ is the simulation output from model (\ref{basic}). The results we obtain are sumarized in Figure \ref{fig:influenza}(a) - \ref{fig:influenza}(b) and strongly suggest that the two outbreaks appear to have shared the same epidemiological characteristics. Figure \ref{fig:influenza}(a) shows the posterior distribution of the four-parameter model. The marginal posterior distributions of $q_{h1}$ and $q_{c1}$ are largely overlapping with the marginal posterior distributions of $q_{h2}$ and $q_{c2}$ and we therefore, unsurprisingly, get strong evidence in favour of the two-parameter model. Figure \ref{fig:influenza}(b) shows the marginal posterior distribution of the model; the posterior probability of model 1 is 0.98 (median over $10$ runs), which gives unambiguous support to model 1, meaning that outbreaks of the same strain share the same dynamics. 
\\
\\
Outbreaks due to a different viral strain (Suppl. Table 3) have different characteristics as indicated by the posterior distribution of the four-parameter model presented in Figure \ref{fig:influenza}(c). This was confirmed by applying our model selection algorithm; the inferred posterior marginal model probability of a two-parameter model was negligible (results not shown). From Figure \ref{fig:influenza}(c) we also see that these differences are due to differences in viral spread across the community whereas within-household dynamics are comparable.  We thus explore a further model with three parameters, $q_{c1}$, $q_{c2}$, $q_{h}$ (model 1), where the two outbreaks share the same within-household characteristics ($q_h$), and compare it against and the four-parameter model (model 2). The obtained Bayes factor suggests that there is only very week evidence in favour of model 1 (Figure \ref{fig:influenza}(d)), which is in agreement with the result of \cite{Clancy:2007p23340}.
\\
\\
In general genetic predisposition, differences in immunity and lifestyle etc. will lead to heterogeneity in susceptibility to viral infection among the host population. Such a model can be written as \cite{ONeill:2000p23341}
\begin{equation}
w_{js}(v) = \sum_{i=0}^{s-j}\binom{s}{i}v^i(1-v)^{s-i}w_{j,s-i}. \label{model_v}
\end{equation}
On the basis of the previous results, we combine both outbreak data sets from Suppl. Table 2, and find some evidence that model (\ref{model_v}) explains the data better than model (\ref{basic}), suggesting that the host-virus dynamics are shaped by the molecular nature of the viral strain, as well as by variability in the host population (see Suppl. Figure 2).

\subsection{JAK-STAT signaling pathway}
\begin{figure}[thp]	\centering
	\includegraphics[width=8cm]{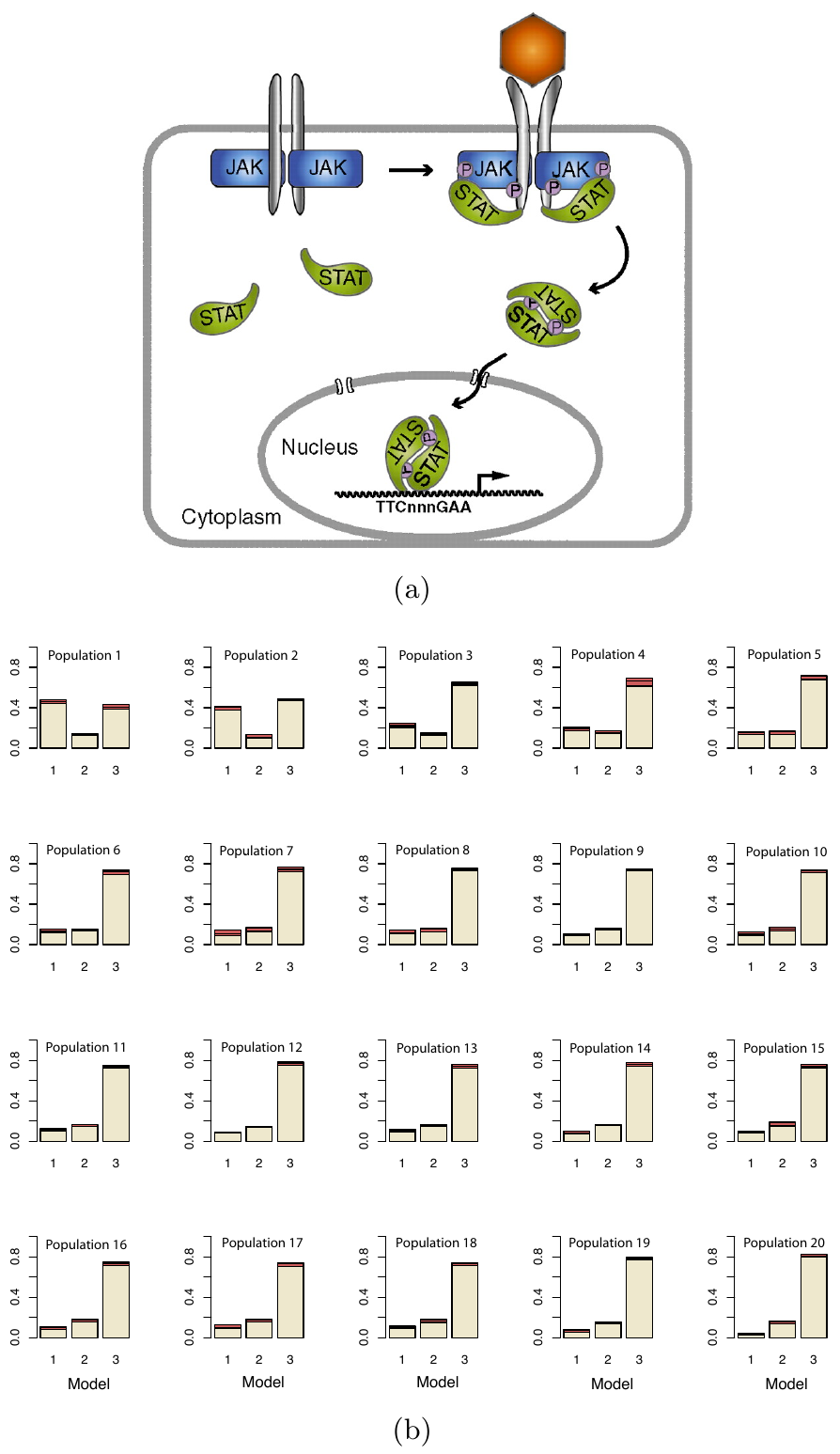}
	\caption{\small{
	(a) STAT5 signaling pathway. Adapted from \cite{Arbouzova:2006p27905}.
	(b) Histograms show populations of the model parameter $m$. Population 20 represents the approximation of the marginal posterior distribution of $m$. Tolerance schedule: $\epsilon = \{$200, 100, 50, 35, 30, 25, 22, 20, 19, 18, 17, 16, 15, 14, 13, 12, 11, 10, 9, $8\}$. Perturbation kernels: $KM_t(m|m^*) = 0.6$ if $m = m^*$ and $0.2$ otherwise; $KP_t(\theta|\theta^*) = \theta^* + U(-\sigma,\sigma)$, $\sigma = 0.5 (\max\{\theta\}_{t-1} -\min\{\theta\}_{t-1})$. $N=500$. Distance function: 
$d(D_0,D^*) = \sqrt{\sum_t{\left( \frac{y_0^{(1)}(t) -y^{*(1)}(t)}{\sigma^{(1)}_{D_0}(t))} \right)^2+ \left( \frac{y^{(2)}(t) - y^{*(2)}(t)}{\sigma^{(2)}_{D_0}(t)} \right)^2 }}$, with $D_0 = \{y_0^{(1)},y_0^{(2)} \}$, $D^*= \{y^{*(1)},y^{*(2)} \}$ and $y^{(1)}$ the total amount of phosphoryalated STAT5 in the cytoplasm and $y^{(2)}$ the total amount of STAT5 in the cytoplasm.  $\sigma_{D_0}^{(1)}$ and $\sigma_{D_0}^{(2)}$ are the associated confidence intervals; reassuringly, other distance functions, e.g. the square root of the sum of squared errors yield identical model selection results (data not shown).
		 \label{fig:stat5} }}
\end{figure}

Having convinced ourselves that the novel ABC SMC model selection approach agrees with the analytical model probabilities, and those obtained using conventional Bayesian model selection, while outperforming conventional ABC rejection model selection approaches, we can now turn our attention to real world scenarios that have not previously been considered from a Bayesian (exact or approximate) perspective. Here we consider models of signaling though the erythropoietin receptor (EpoR), transduced by STAT5 (Figure \ref{fig:stat5}(a)) \cite{Darnell:1997p2005, Horvath:2000p1959}. Signaling through this receptor is crucial for proliferation, differentiation, and survival of erythroid progenitor cells \cite{Klingmuller:1996p1955}. 
When the Epo hormone binds to the EpoR receptor, the receptor's cytoplasmic domain is phosporylated, which creates a docking site for signaling molecules, in particular  STAT5. Upon binding to the activated receptor, STAT5 first becomes phosphorylated, then dimerizes and translocates to the nucleus, where it acts as a transcription factor. There have been competing hypotheses about what happens with the STAT5 in the nucleus. Originally it had been suggested that STAT5 gets degraded in the nucleus in an ubiquitin-asssociated way \cite{Kim:1996p1953}, but other evidence suggests that they are dephosphorylated in the nucleus and then trafficked back to the cytoplasm \cite{Koster:1999p1954}. 
\\
\\
The ambiguity of the shutoff mechanism of  STAT5 in the nucleus triggered the development of several mathematical models \cite{Swameye:2003p130, Muller:2004p265, Timmer:2004p474} describing different hypotheses. All models assume mass action kinetics and denote the amount of activated Epo-receptors by $EpoR_A$, monomeric unphosphorylated and phosporylated STAT5 molecules by $x_1$ and $x_2$, respectively, dimeric phosphorylated STAT5 in the cytoplasm by $x_3$ and dimeric phosphorylated STAT5 in the nucleus by $x_4$. 
The most basic model Timmer \textit{et al.} developed, under the assumption that phosphorylated STAT5 does not leave the nucleus, consists of the following kinetic equations,
\begin{eqnarray}
	\dot{x}_1 &=& - k_1 x_1 EpoR_A \label{eq:basic_1} \label{first} \\
	\dot{x}_2 &=& - k_2 x_2^2 + k_1 x_1 EpoR_A \nonumber \\
	\dot{x}_3 &=& - k_3 x_3 + \frac{1}{2}k_2x_2^2  \nonumber \\
	\dot{x}_4 &=& k_3 x_3  \label{eq:basic_4}. \label{fourth}
\end{eqnarray}
One can then assume that phosphorylated STAT5 dimers dissociate and leave the nucleus; this is modelled by adding appropriate kinetic terms to the equations (\ref{first}) and (\ref{fourth}) of the basic model to obtain
\begin{eqnarray*}
	\dot{x}_1 &=& - k_1 x_1 EpoR_A + 2 k_4 x_4\\
	\dot{x}_4 &=& k_3 x_3 - k_4 x_4.
\end{eqnarray*}
The cycling model can be developed further by assuming a delay before STAT5 leaves the nucleus:
\begin{eqnarray}
	\dot{x}_1 &=& - k_1 x_1 EpoR_A + 2 k_4 x_3(t-\tau) \nonumber \\
	\dot{x}_4 &=& k_3 x_3 - k_4 x_3(t-\tau) \label{strange}.
\end{eqnarray}
This model was chosen as the best model in the original analyses \cite{Muller:2004p265, Swameye:2003p130} based on a numerical evaluation of the likelihood, followed by a likelihood ratio test and bootstrap procedure for model selection. The data are partially observed time course measurements of the total amount of STAT5 in the cytoplasm, and the amount of phosphorylated STAT5 in the cytoplasm; both are only known up to a normalizing factor.
\\
\\
We propose a further model with clear physical interpretation where the delay acts on STAT5 inside the nucleus ($x_4$) rather than on $x_3$ (in equation (\ref{strange})), for which a biological interpretation is difficult. Instead of $x_3(t-\tau)$, we propose to model the delay of phosphorylated STAT5 $x_4$ in the nucleus directly and obtain \cite{Zi:2006}:
\begin{eqnarray*}
	\dot{x}_1 &=& - k_1 x_1 EpoR_A + 2 k_4 x_4(t-\tau)\\
	\dot{x}_4 &=& k_3 x_3 - k_4 x_4(t-\tau).
\end{eqnarray*}
We perform the ABC SMC model selection algorithm on the following non-nested models: (1) Cycling delay model with $x_3(t-\tau)$, (2) Cycling delay model with $x_4(t-\tau)$, (3) Cycling model without a delay. The model parameter $m$ can therefore take values 1, 2 and 3. 
\\
\\
For each proposed model and parameter combination we numerically solve the ODE equations of the model and add $\epsilon \sim N(0,\sigma)$ to obtain the simulated time course data. The noise parameter $\sigma$ can be either fixed or treated as another parameter to be estimated; we consider the latter option, under the assumption that the experimental noise is independent and identically distributed for all time points. 
\\
\\
Figure \ref{fig:stat5}(b) shows intermediate populations leading to the ABC SMC marginal posterior distribution over the model parameters $m$ (population 20). Bayes factors can be calculated from the last population and according to the conventional interpretation of Bayes factors \cite{Kass:1995p2898}, it can be concluded that there is strong evidence in favour of model 3 compared to model 1, positive evidence in favour of model 3 compared to model 2, and positive evidence in favour of model 2 compared to model 1. Thus cycling appears to be clearly important and the model that receives the most support is the cycling model without a time-delay. Here the flexibility of ABC SMC has allowed us to perform simultaneous model selection on non-nested models of ordinary and time-delay differential equations.

\section{Discussion}

We have developed a novel model selection methodology based on approximate Bayesian computation and sequential Monte Carlo. The results obtained in our  applications illustrate the usefulness and wide applicability of our ABC SMC method, even when experimental data are scarce, there are no measurements for some of the species, temporal data are not measured at equidistant time points, and when parameters such as kinetic rates are unknown.  In the context of dynamical systems our method can be applied across all simulation and modelling (including qualitative modelling) frameworks; for JAK-STAT signal transduction dynamics, for example, we have been able to compare the relative explanatory power of ODE and time-delay differential equation models. Our model selection procedure is also not confined to dynamical systems; in fact the scope for application is immense and limited only by the availability of efficient simulation approaches. 
\\\\
Routine application to complex models in systems, computational and population biology 
with hundreds or thousands of parameters \cite{Chen:2009p21129} will require further numerical developments due to the high computational cost of repeated simulations. SMC based ABC methods are, however, highly paralellizable and we believe that future work should exploit this property to make these methods computationally more efficient. Further potential improvements might come from (i) regression adjustment techniques that have so far been applied in the parameter estimation ABC framework \cite{Beaumont:2002p13862, Blum:2009p24290,Excoffier:2009p20927} (ii) from automatic generation of the tolerance schedules \cite{Jasra}, and (iii) by developing more sophisticated perturbation kernels that exploit inherent properties of biological dynamical systems such as sloppiness \cite{Gutenkunst:2007p728,Secrier:2009}; here especially we feel that there is substantial room for improvement as the likelihoods of dynamical systems contain information about the qualitative behaviour \cite{Kirk:2008p20329} which can also be exploited in ABC frameworks. 

\section{Conclusion}

We conclude by emphasizing the need for inferential methods which can assess the relative performance and reliability of different models. The need for such reliable model selection procedures can hardly be overstated: with an increasing number of biomedical problems 
being studied using simulation approaches, there is an obvious and urgent need for statistically sound approaches that allow us to differentiate between different models. If parameters are known or the likelihood is available in a closed form, then the model selection is generally straightforward. However, for many of the most interesting systems biology (and generally, scientific) problems this is not the case and here ABC SMC can be employed. 

\section*{Acknowledgement}
We are especially grateful to Paul Kirk for his insightful comments and many valuable discussions. We furthermore thank the members of Theoretical Systems Biology Group for discussions and comments on earlier versions of this paper.\\
\\
\textit{Funding:} This work was supported through a MRC priority studentship (T.T.) and BBSRC grant BB/G009374/1.

\bibliographystyle{references.bst} 
\bibliography{ABC_ms_ref}

\newpage

\section*{Supplementary material A: Derivation of ABC SMC model selection algorithms}

We start this section by briefly reviewing the building blocks of the ABC SMC algorithm of Toni \textit{et al.}\cite{Toni:2009p20998}, which is based on sequential importance sampling (SIS). The main idea of importance sampling is to sample from the desired target distribution $\pi$ (which can be impossible or hard to sample from) indirectly through sampling from a proposal distribution $\eta$ \cite{PRobert:2004p16815}. To get a sample from $\pi$,  one can instead sample from $\eta$ and weight the samples by importance weights
$$
w(x) = \frac{\pi(x)}{\eta(x)}.
$$
In SIS one reaches the target distribution $\pi_T$ through a series of intermediate distributions,  $\pi_t$, $t = 1,\ldots,T-1$ \cite{Doucet:2001p27492, DelMoral:2006p4}. If it is hard to sample from these distributions one can use the idea of importance sampling described above to sample from a series of proposal distributions $\eta_t$ and weight the obtained samples by importance weights 
\begin{equation} \label{seq_weights} 
w_t(x_t) = \frac{\pi_t(x_t)}{\eta_t(x_t)}. 
\end{equation}
In SIS the proposal distributions are defined as
\begin{equation} \label{eq:proposal_distributions}
\eta_t(x_t) = 
\int \eta_{t-1}(x_{t-1}) \kappa_t(x_{t-1},x_t) dx_{t-1},
\end{equation}
where $ \eta_{t-1}$ is the previous proposal distribution and $\kappa_t$ is a Markov kernel. \\
\\
To apply SIS, we need to define the intermediate and the proposal distributions. In an ABC framework \cite{Beaumont:2002p13862, Marjoram:2003p5}, which is based on comparisons between simulated and experimental datasets, we define the intermediate distributions as \cite{Sisson:2007p2, Toni:2009p20998}
\begin{equation}
\pi_t(x) = \frac{P(x)}{B_t}\sum_{b=1}^{B_t} \mathbb{1} \left( d(D_0,D_{(b)}(x)) \leq \epsilon_t \right), \label{intermediate_dist_msel}
\end{equation}
where $P(x)$ denotes the prior distribution and $D_{(1)}, \ldots, D_{(B_t)}$ are $B_t \geq 1$ data sets generated for a fixed parameter $x$,
$D_{(b)} \sim f(D|x)$. $\mathbb{1}(x)$ is an indicator function and  $\epsilon_t$ is the tolerance required from particles contributing to the intermediate distribution  $t$. To simplify the notation we define $b_t(x) = \frac{1}{B_t}\sum_{b=1}^{B_t} \mathbb{1}\left( d(D_0,D_{(b)}(x)) \leq \epsilon_t \right)$. 
\\
\\
We define the first proposal distribution to equal the prior distribution, $\eta_1(x) = P(x)$. The proposal distribution at time $t$ ($t=2,\ldots,T$), $\eta_t$, is defined as
\begin{eqnarray}
\eta_t(x_t) &=& \mathbb{1} \left( P (x_t)>0 \right) \mathbb{1} \left( b_t(x_t)>0 \right) \int \pi_{t-1}(x_{t-1}) K_t(x_t|x_{t-1})  dx_{t-1}  \label{prop_abc},
\end{eqnarray}
where  $K_t$ denotes the perturbation kernel (e.g. random walk around the particle). For details of how this proposal distribution was obtained, see \cite{Toni:2009p20998}. \\
\\
In the remainder of this section we introduce three different ways in which ABC SMC ideas presented above can be used in the model selection framework. We start by proposing a simple and naive incorporation of the above building blocks for model selection. We then continue by deriving an ABC SMC model selection algorithm on the joint model and parameter space, which is presented in the methods section of the paper. In the end we present ABC SMC algorithm for approximation of the marginal likelihood, which can also be employed for model selection. \\
\\
The only of these three algorithms that we present in the main part of the paper and use in examples is algorithm II (ABC SMC model selection on the joint space), since the other two algorithms (I and III) are computationally too expensive and impractical to use.
\\
\\
I) \textit{ABC SMC$_m$ REJ$_{\theta}$ model selection algorithm}\\
\\ 
Very naively and stragihtforwardly the intermediate distributions can be defined as
$$
\pi_t(m) =P(m) bm_t(m),
$$
where
$$
bm_t(m):=  \frac{\sum_{\theta \sim P(\theta|m)}\mathbb{1}(d(D_0, D(\theta,m))< \epsilon_t)}{\sum_{\theta \sim P(\theta|m)}\mathbb{1}(P(\theta|m) > 0)}.
$$
This means that for each model $m$ we calculate $bm_t(m)$ as the ratio between the number of accepted particles (where the distance falls below $\epsilon_t$) and all sampled particles, where parameters $\theta$ of model $m$ are sampled from the prior distribution $P(\theta|m)$.
\\
\\
If a set of candidate models $\mathcal{M}$ of a finite size $|\mathcal{M}|$ is being considered, and $N$ denotes the number of particles, then we can write the algorithm as follows:
\begin{description}

 	\item [MS1] Initialize $\epsilon_1, \ldots, \epsilon_T$.\\
	Set the population indicator $t = 1$. 

	\item [MS2]  For $i=1,\ldots,|\mathcal{M}|$, calculate the weights as
	
	$$
	w_t^{(i)}(m_t^{(i)} ) =
	\left\{ \begin{array}{ll}
	bm_t(m_t^{(i)}),  &  \textrm{if } t=1  \\
	\frac{P(m_t^{(i)})bm_t(m_t^{(i)})}{ \sum_{j=1}^N w_{t-1}^{(j)} KM_t(m_t^{(i)}|m_{t-1}^{(j)})}, & \textrm{if } t >1.
	\end{array} \right.
	$$

	\item [MS3] Normalize the weights.\\
	If $t<T$, set $t = t+1$, go to \textbf{MS2}.   

\end{description}
In this algorithm we estimate the posterior distribution of the model indicator $m$ sequentially (i.e. using ideas from SIS), but the integration over model parameters is not sequential; we always sample them from the prior distribution $P(\theta|m)$ (i.e. in the rejection sampling manner). This algorithm is therefore computationally very expensive. It would be computationally more efficient to generate $\theta_t$ by exploiting the knowledge about $\theta$ that is contained in $\{ \theta \}_{t-1}$. In addition to learning $m$ sequentially, i.e. by exploiting $\{m\}_{t-1}$ for generating $m_t$, we would also like to learn $\theta$ sequentially.\\
\\
In order to do this, we define \\
\\
II) \textit{ABC SMC model selection on the joint space}\\
\\
Let $(m,\theta)$ denote a particle from a joint space, where $m$ corresponds to the model indicator and $\theta$ are the parameters of model $m$. We define the intermediate distributions by 
$$
\pi_t(m,\theta) = P(m,\theta) b_t(m,\theta),
$$
where
$$
b_t(m,\theta) = \frac{1}{B_t} \sum_{b=1}^{B_t} \mathbb{1}(d(D_0,D_{(b)}(m,\theta)) \leq \epsilon_t).
$$
In the following equations $KM_t$ denotes the perturbation kernel for the model parameter, $KP_{t,m}$ denotes the perturbation kernel for the parameters of model $m$, and $t$ is the population number. Now we derive the sequential importance sampling weights
$$
w_t(m_t,\theta_t) = \frac{\pi_t(m_t,\theta_t)}{\eta_t(m_t,\theta_t)}.
$$
For a particle $(m_t,\theta_t)$ from population $t$, we define the proposal distribution $\eta_t(m_t,\theta_t)$ as
\begin{eqnarray}
\eta_t(m_t,\theta_t)&=& 1 \left( P(m_t,\theta_t)>0 \right) 1 \left( b_t(m_t,\theta_t)>0 \right)  \label{prop_distributions_msel} \\
&& \times \int_{m_{t-1}}  \pi_{t-1}(m_{t-1})  KM_t (m_t|m_{t-1}) dm_{t-1} \nonumber \\
&& \times \int_{\theta_{t-1}|m_{t-1} = m_t}  \pi_{t-1}(\theta_{t-1}) KP_t(\theta_t|\theta_{t-1}) d\theta_{t-1} \nonumber \\
&\propto& 1 \left( P(m_t,\theta_t)>0 \right) 1 \left( b_t(m_t,\theta_t)>0 \right) \nonumber \\
&& \times \sum_{j=1}^{|\mathcal{M}|} P_{t-1}(m_{t-1}^{(j)})  KM_t (m_t|m_{t-1}^{(j)}) \nonumber \\
&& \times \sum_{k;m_{t-1}=m_t} \frac{w_{t-1}^{(k)}}{P_{t-1}(m_{t-1}=m_t)} KP_{t,m_t} (\theta_t|\theta_{t-1}^{(k)}), \nonumber
\end{eqnarray}
where intermediate marginal model probabilities $P_t(m)$ are defined as
$$
P_t(m_t=m) = \sum_{m_t=m} w_t(m_t,\theta_t).
$$
The weights for all \textit{accepted} particles are (obtained by including (\ref{intermediate_dist_msel}) and (\ref{prop_distributions_msel}) in equation (\ref{seq_weights}))
$$
w_t(m_t,\theta_t) = \frac
{P(m_t,\theta_t) b_t(m_t,\theta_t) }
{
\sum_{j=1}^{|\mathcal{M}|} P_{t-1}^{(j)}(m_{t-1}^{(j)}) KM_t (m_t|m_{t-1}^{(j)}) \sum_{k;m_{t-1}=m_t} \frac{w_{t-1}^{(k)}}{P_{t-1}(m_{t-1}=m_t)} KP_{t,m_t} (\theta_t|\theta_{t-1}^{(k)})
}.
$$
The resulting ABC SMC algorithm is presented in the methodology section of the main part of the paper.\\
\\
III) \textit{ABC SMC approximation of the marginal likelihood $P(D_0|m)$}\\
\\
If we can calculate the marginal likelihood $P(D_0|m)$ for each of the candidate models that we consider in the model selection problem, then we can calculate the marginal posterior distribution of a model $m$ as
\begin{equation}
P(m|D_0) = \frac{P(D_0|m)P(m)}{\sum_{m'}P(D_0|m')P(m')}. \label{marg_lik}
\end{equation}
We now explain how to calculate $P(D_0|m)$ for model $m$. In the ABC rejection-based approach the posterior distribution of the parameters for each model $m$ are estimated independently by employing ABC rejection; the marginal likelihood then equals the acceptance rate, 
\begin{equation}
P(D_0|m) \approx \frac{\#\textrm{accepted particles given model m}}{N_m}, \label{marg_likelihood}
\end{equation}
i.e. the ratio between the number of accepted versus the number of \textit{proposed} particles $N_m$. We can use this marginal likelihood estimate to calculate $P(m|D_0)$ using equation (\ref{marg_lik}). This approach has been used in \cite{Wilkinson:2007p14173}.\\
\\
We now derive how ABC SMC can be used for estimating the marginal likelihood, which can be then used for model selection. In a usual ABC SMC setting for drawing samples from the posterior parameter distribution $P(\theta|m,D_0)$ for a given model $m$, we define intermediate distributions as 
\begin{equation}
\pi_t(\theta) = P(\theta)\mathbb{1}(d(D_0, D(\theta))\leq \epsilon_t). \label{interm_pops}
\end{equation}
The target distribution $\pi_T$ is an unnormalized approximation of the posterior distribution $P(\theta|m,D_0)$. We are now interested in its normalization constant, i.e. the marginal likelihood,
$$
P(D_0|m) \approx \int_{\theta} \pi_T(\theta) d\theta.
$$
Let us call the integrals of $\pi_t(\theta)$, $\int_{\theta}\pi_t(\theta) d\theta$, the \textit{intermediate marginal likelihoods}.\\
\\
In the usual ABC SMC parameter estimation setting, our goal is to obtain samples from distribution $\pi_T(\theta)$, whereas our goal here is to obtain its normalization constant. While this distribution as defined in equation (\ref{interm_pops}) is in general unnormalized, the ABC SMC parameter estimation algorithm performs normalization of weights at every $t$ and therefore returns its normalized version\cite{Toni:2009p20998}. So we cannot use the usual output of ABC SMC directly. Instead we proceed as follows. \\
\\
We would like to draw particles from the following target distribution:
$$
\mathcal{T}_T(\theta) = P(\theta)\mathbb{1}[d (D_0, D(\theta))\leq \epsilon_T] + P(\theta)\mathbb{1}[d (D_0, D(\theta)) >\epsilon_T],
$$
where $P(\theta)$ is the prior distribution. To draw samples from $\mathcal{T}_T$ we can use ABC SMC, where we define the intermediate distributions as
\begin{eqnarray*}
\mathcal{T}_t(\theta) &=& P(\theta)\mathbb{1}[d (D_0, D(\theta)) \leq \epsilon_t] + P(\theta)\mathbb{1}[d (D_0, D(\theta)) >\epsilon_t] \\
&=& \mathcal{T}_t^{1}(\theta) + \mathcal{T}_t^{2}(\theta) .
\end{eqnarray*}
In each population we accept $N$ particles, and a particle is only rejected if it falls outside the boundaries of $\mathcal{T}_t$. We classify the accepted particles in two sets, $\Theta^1_t := \{ \theta; d (D_0, D(\theta))  \leq \epsilon_t\}$ and $\Theta^2_t := \{ \theta; d (D_0, D(\theta))  > \epsilon_t\}$, depending on the distance reached. In each population $t$ we can then calculate the intermediate marginal likelihoods by
$$
\int_{\theta}\mathcal{T}_t^{1}(\theta) d\theta = \sum_{\theta \in \Theta^1_t} w_t(\theta).
$$
The target marginal likelihood, $\int_{\theta}\mathcal{T}_T^{1}(\theta) d\theta$,  is our approximation of $P(D_0|m)$. In an ABC rejection setting, where $T=1$ and all weights are equal, this result corresponds to (\ref{marg_likelihood}).\\
\\
After calculating $P(D_0|m)$ for each $m$, we can use equation (\ref{marg_lik}) to calculate the marginal posterior distributions for model $m$,
$$
P(m|D_0) \approx \frac{P(m) \sum_{\theta \in \Theta_T^1} w_T(\theta)}{ P(m')\sum_{m'} \sum_{\theta' \in \Theta'^1_T} w'_T(\theta')}.
$$ 
The model selection algorithm based on approximating the marginal likelihood proceeds as follows:

\subsection*{Algorithm}

\begin{description}

	\item [M1] For model $m_j$, $j=1,\ldots,|\mathcal{M}|$ do steps S1 to S4. Then go to \textbf{M2}.
	
 	\item [S1] Initialize $\epsilon_1, \ldots, \epsilon_T$.\\
	Set the population indicator $t = 1$. 

	\item [S2.0] Set the particle indicator $i=1$.
	
	\item [S2.1] If $t = 1$, sample $\theta^{**}$ independently from $P(\theta)$.\\
	If $t > 1$, sample $\theta^*$ from the previous population $\{\theta_{t-1}^{(i)}\}$ with weights $w_{t-1}$ and perturb the particle to obtain $ \theta^{**} \sim K_t(\theta|\theta^{*})$, where $K_t$ is a perturbation kernel.\\ 
	If $P(\theta^{**}) = 0$, return to \textbf{S2.1}.\\
	For a particle $\theta^{**}$ simulate a candidate data set $D$ and calculate $d(D_0,D(\theta^{**}))$.\\
	If $d(D_0,D(\theta^{**})) \leq \epsilon_t$, add $\theta^{**}$ to $\Theta_t^1 (m_j)$. If $d(D_0,D(\theta^{**})) > \epsilon_t$, add $\theta^{**}$ to $\Theta_t^2(m_j)$. 
		
	\item [S2.1] Calculate the weight for particle $\theta_t^{(i)} = \theta^{**}$:
	$$
	w_t^{(i)}(\theta_t^{(i)}) =
	\left\{ \begin{array}{ll}
	1,  &  \textrm{if } t=1  \\
	P(\theta_t^{(i)})/ \left( \sum_{j=1}^{N} w_{t-1}^{(j)} K_t(\theta_t^{(i)}|\theta_{t-1}^{(j)}) \right), & \textrm{if } t >1.
	\end{array} \right.
	$$
	If $i<N$ set $i = i+1$, go to \textbf{S2.1}.
	
	\item [S3] Normalize the weights.\\
	If $t<T$, set $t = t+1$, go to \textbf{S2.0}.   
	
	\item [S4] Calculate
	$$
	P(D_0|m_j) \approx \frac{ P(m_j)\sum_{\theta \in \Theta_T^{1}(m_j)} w_T(\theta)}{ \sum_{m'} P(m') \sum_{\theta' \in \Theta'^1_T(m)} w'_T(\theta')}.
	$$

	\item [M2] For each $m_j$ calculate $P(m_j|D_0)$ using equation	
	$$
	P(m|D_0) = \frac{P(D_0|m)P(m)}{\sum_{m'}P(D_0|m')P(m')}.
	$$ 
	
\end{description}
The computational advantage of this model selection algorithm compared to the marginal likelihood model selection based on ABC rejection can be obtained by (i) starting with a small number of particles $N$ in population $1$ and increasing it in each subsequent population. This way not much computational effort is spent on simulations in earlier populations, but we nevertheless have a big enough sample set in the last population to obtain a reliable estimate; (ii) exploiting the property that intermediate distributions in the parameter estimation framework should be included in one another, and so
$$
\textrm{range }  \Theta^1_{t} \geq \textrm{range }  \Theta^1_{t+1}, \hspace{0.5cm} t=1,\ldots, T-1 .
$$
In other words, a proposed particle in population $t$ cannot belong  to $\Theta^1_{t}$ if it cannot be obtained by perturbing any of the particles in $\Theta^{1}_{t-1}$. We can therefore reject some of the proposed particles without simulation. This means a huge saving in computational time, since simulations are the most expensive part of ABC based algortihms. However, one of the obvious ways to exploit this property would be to use a truncated perturbation kernel with ranges they cover being smaller than the range of prior distribution. But we find this unsatisfactory and, in the present form, feel that evaluating the marginal model likelihood directly is not practical.

\section*{Supplementary material B: Tutorial on ABC rejection and ABC SMC for parameter estimation and model selection}

Available in arXiv (reference arXiv:0910.4472v2 [stat.CO]).

\section*{Supplementary material C: Supplementary figures and datasets}

Available on the Bioinformatics webpage.

\end{document}